\documentstyle[11pt,epsf]{article}
\begin{document}
\renewcommand{\thefootnote}{\fnsymbol{footnote}}
\begin{center}
{\large \bf Polarized Parton Distributions: \\
Theory and Experiments} \\
\vspace{5mm}
Gordon P. Ramsey$^{1,2}$
\footnote{Work supported in part by the U.S. Department of Energy, Division of
High Energy Physics, Contract W-31-109-ENG-38. Presented at the
Circum-Pan-Pacific Workshop on High Energy Spin Physics 96, Kobe, Japan, $2-4$
October, 1996.} \\
\vspace{5mm}
{\small\it
(1) Loyola University of Chicago, Chicago, IL 60626, USA \\
(2) Argonne National Laboratory, IL 60439, USA}
\end{center}

\renewcommand{\thefootnote}{\arabic{footnote}}
                                  
\begin{abstract}
We have extracted the constituent contributions to the spin of the proton from
recent data at CERN, SLAC and DESY. The valence, sea quark and antiquark
spin-weighted distributions are determined separately. The data appear to imply
a small to moderate polarized gluon distribution, so that the anomaly term is
not significant in determining these contributions. We have analyzed the
consistency of the results obtained from various sets of data and the Bjorken
Sum Rule. All data are consistent with the sum rule, but they differ in the
contribution of the strange sea to proton spin. This and the remaining
uncertainty in the polarized gluon distribution pose unanswered questions
about hadronic spin. Only further experiments which extract information about
the polarized gluon and sea will reconcile these differences. We suggest
specific experiments which can be performed to determine the size of the
polarized sea and gluons.
\end{abstract}

\renewcommand{\thefootnote}{\arabic{footnote}}

\begin{center}  \Large
{\bf Introduction}
\end{center}   \normalsize

One of the goals of high energy spin physics is to determine the contributions
of quarks and gluons, as well as the effect of the orbital motion, to nucleon
spin. Significant interest in high energy polarization was generated when the
European Muon Collaboration (EMC)$^{1}$ analyzed polarized deep-inelastic
lepton-hadron scattering (DIS) data, which implied that the Bjorken sum rule
(BSR) of QCD$^{2}$ was satisfied and the Ellis-Jaffe sum rule$^{3}$ based on
a simple quark model was violated. Recently, the Spin Muon Collaboration (SMC)
group from CERN$^{4}$, the experimental groups from SLAC$^{5}$ and the HERMES
group at DESY$^{6}$ measured the polarized structure functions $g_1^p$ to low
$x$ and have added the corresponding neutron and deuteron structure functions
$g_1^n$ and $g_1^d$. They also improved statistics and lowered the systematic
errors from the EMC data.

An advantage to using spin in probing hadronic structure is that the
theoretically calculated spin-weighted parton distributions are related to
the experimentally measured cross sections by the polarized structure
functions, $g_1$. The measured asymmetry in deep-inelastic lepton-hadron
scattering (DIS) is given by:
\begin{equation}
A=\Biggl[{{\sigma(\rightarrow \rightarrow)-\sigma(\rightarrow \leftarrow)}
\over {\sigma(\rightarrow \rightarrow)+\sigma(\rightarrow \leftarrow)}}
\Biggr]  \label{1}
\end{equation}
$$
\quad =D(A_1+\eta A_2).
$$
The spins of the target and beam are aligned or anti-aligned with the momentum
of the intraction in measuring these cross sections. The terms $D$ and $\eta$
are known kinematic factors for each experiment. The structure function $g_1$
is normally extracted from the asymmetry $A_1$ by using the approximation
\begin{equation}
g_1^p(x,Q^2)\approx {{A_1(x)\>F_2(x,Q^2)}\over {2x(1+R)}}, \label{2}
\end{equation}
where $F_2$ is the corresponding unpolarized singlet structure function and
$R=\surd (\sigma_L/\sigma_T)$, the ratio of the cross sections for absorbing
longitudinal and transverse virtual photons. It is assumed that the transverse
part of the asymmetry $A_2^p$ is small and that $A_1$ is relatively independent
of $Q^2$, which has been implied by these experiments.

The integrated polarized structure function, $I^{p(n)}\equiv
\int_0^1 g_1^{p(n)}(x)\>dx$, is related to the polarized quark distributions by
$$
I^{p(n)}={1\over {18}}(1-\alpha_s^{corr})\langle\bigl[4(1)\Delta u_v+1(4)
\Delta d_v+4(1)(\Delta u_s+\Delta \bar{u})+1(4)(\Delta d_s+\Delta \bar{d})
$$
\begin{equation}
\qquad \qquad \qquad +(\Delta s+\Delta \bar{s})\bigr]\rangle. \label{3}
\end{equation}                                         
The QCD corrections, characterized by $\alpha_s^{corr}$, have been calculated
to $O(\alpha_s^4)$$^{7,8}$ and are
\begin{equation}
\alpha_s^{corr}\approx ({{\alpha_s}\over {\pi}})+3.5833({{\alpha_s}\over
{\pi}})^2+20.2153({{\alpha_s}\over {\pi}})^3+130({{\alpha_s}\over {\pi}})^4,
\label{4}
\end{equation}                                         
where the last term is estimated. Equations (1) through (4) provide a direct
means to extract information about the polarized quark distributions from the
DIS experiments. There have been some recent theoretical approaches to this
problem.$^{7,9}$ We have done a detailed flavor dependent analysis including
the QCD corrections and the effect of the gluon anomaly. It is assumed that
the polarized gluon distribution is of small to moderate size and we determine
the resulting polarized quark distributions for each set of data using the
appropriate sum rules. The key elements of our approach are:
\begin{itemize}
\item
determine the valence contribution to the spin using the BSR
\item 
find sea integrated parton distributions for each flavor by breaking the
SU(6) symmetry with the strange quarks and using the sum rules with data
as input
\item
include higher order QCD corrections and the gluon anomaly for each flavor
\item
discuss similarities and differences between the phenomenological implications
of the different experimental results, and
\item
suggesting a set of experiments which would distinguish the quark and gluon
contributions to the proton spin.
\end{itemize}

This approach differs from that of others in that we use sum rules in
conjunction with a single experimental result to extract the spin information
and we break the flavor symmetric sea while including anomaly contributions.

\begin{center}  \Large
{\bf Theoretical Background}
\end{center}   \normalsize

{\bf Valence Quarks} \\

Fundamentally, we assume that the proton is comprised of valence quarks,
whose integrated polarized distribution is given by: $\langle \Delta q_v (Q^2)
\rangle$. We construct the polarized valence quark distributions from the
unpolarized ones by starting with a 3-quark model based on an SU(6)
proton wave function. The valence quark distributions can be written as:
$$\Delta u_v (x,Q^2)=\cos \theta_D [u_v(x,Q^2)-{2\over 3}d_v(x,Q^2)],$$
\begin{equation}
\Delta d_v (x,Q^2)=-{1\over 3}\cos \theta_D d_v(x,Q^2), \label{5}
\end{equation}                                        
where $\cos \theta_D$ is a ``spin dilution" factor which vanishes as $x\to 0$
and becomes unity as $x\to 1$, characterizing the valence quark helicity
contribution to the proton.$^{10,11}$ The spin dilution factor is adjusted
to satisfy the Bjorken Sum Rule (BSR). The BSR relates the polarized structure
function $g_1(x)$, measured in polarized deep-inelastic scattering, to the
axial vector current $A_3$, which is measured in neutron beta decay. This sum
rule is considered to be a fundamental test of QCD. In terms of the polarized
distributions and our assumptions about the flavor symmetry of the
$u$ and $d$ polarized sea, the BSR can be reduced to:
\begin{equation}
\int_0^1 [\Delta u_v(x,Q^2)-\Delta d_v(x,Q^2)]\>dx=A_3(1-{{\alpha_s}\over
\pi}+\dots). \label{6}
\end{equation}                                         
Thus, the valence contributions are determined uniquely by this model. 
The valence distributions are not sensitive to the unpolarized distributions
used to generate them.$^{12,13}$ With our values $\langle
\Delta u_v\rangle=1.00\pm 0.01$ and $\langle \Delta d_v\rangle=-.26\pm 0.01$,
both the BSR and magnetic moment ratio, $\mu_p/\mu_n\approx -3/2$ are
satisfied. This results in a spin contribution from the valence quarks equal
to $0.74\pm 0.02$. The quoted errors arise from data on $A_3=g_A/g_V$, and the
differences in choice of the unpolarized distributions. \\

{\bf Sea Quarks} \\

The proton is also filled with a quark sea, whose lightest flavors should
dominate the spin, since the heavier quarks would be significantly harder to
polarize. We assume that the quark and antiquark flavors are symmetric, but
break the SU(6) symmetry of the sea by assuming that the polarization of the
heavier strange quarks is suppressed.$^{10}$ The sea distributions are then
related by:
$$\Delta \bar{u}(x,Q^2)=\Delta u(x,Q^2)=\Delta \bar{d}(x,Q^2)=\Delta d(x,Q^2)$$
\begin{equation}
\qquad \qquad =[1+\epsilon]\Delta \bar{s}(x,Q^2)=[1+\epsilon]\Delta s(x,Q^2).
\label{7}
\end{equation}                                         
The $\epsilon$ factor is a measure of the increased difficulty in polarizing
the strange quarks. The DIS data are used to determine $\epsilon$ and the
overall size of the polarized sea. Additional constraints are provided by the
axial-vector current operators, $A_8$ and $A_0$.

The coefficient $A_8$ is determined by hyperon decay and is related to the
polarized quark distributions by:
\begin{equation}
A_8=\langle\bigl[\Delta u_v+\Delta d_v+\Delta u_s+\Delta \bar{u}+\Delta d_s+
\Delta\bar{d}-2\Delta s-2\Delta \bar{s}\bigr]\rangle \approx 0.58\pm 0.02.
\label{8}
\end{equation}
$A_0$ is related to the total spin carried by the quarks in the proton.
We can relate these axial currents and the structure function $g_1^p$ in the
anomaly-independent form:
\begin{equation}
A_0= 9(1-\alpha_s^{corr})^{-1}\int_0^1 g_1^p(x)\>dx-{1\over 4}A_8-
{3\over 4} A_3\approx \langle \Delta q_{tot}\rangle. \label{9}
\end{equation}

{\bf Gluons} \\

The gluons are polarized through Bremsstrahlung from the quarks. The integrated
polarized gluon distribution is written as: $\langle\Delta G\rangle=\int_0^1
\Delta G(x,Q^2)\>dx$. We cannot determine {\it a priori} the size of the
polarized gluon distribution at a given $Q^2$. The evolution equations for the
polarized distributions indicate that the polarized gluon distribution
increases with $Q^2$ and that its evolution is directly related to the behavior
of the orbital angular momentum.$^{14}$ Thus, one assumes a particular form
for the polarized gluon distribution at a given $Q^2$ and checks its
consistency with data which are sensitive to $\Delta G(x,Q^2)$ at a particular
$Q_0^2$.

The model of $\Delta G$ that is used has a direct effect on the measured value
of the quark distributions through the gluon axial anomaly,$^{15}$ which has
the form:
\begin{equation}
\Gamma (Q^2)={{N_f\alpha_s(Q^2)}\over {2\pi}}\int_0^1 \Delta G(x,Q^2)\>dx,
\label{10}
\end{equation}                                         
where $N_f$ is the number of quark flavors. For each quark flavor, the measured
polarization distribution is modified by a factor: $\langle\Delta q_i\rangle-
\Gamma(Q^2)/N_f$. Thus, the quark spin contributions depend indirectly on the
polarized gluon distribution. In a naive quark model, $\langle \Delta q
\rangle=1$ and $\Delta G$ may be quite large to be consistent with data.$^{16}$
However, a reasonably sized $\Delta G$ is possible if the sea has a suitably
negative polarization. We have considered two possible models for
calculating the anomaly: (1) $\Delta G=xG$ (indicating that the spin carried
by gluon is equal to its momentum) and (2) $\Delta G=0$, which is an extreme
case for bounding the distributions. We believe that the present data imply
that anomaly effects, and thus the overall integrated polarized gluon
distribution, is limited at these energies.

The polarized distributions are related to the orbital angular momentum of the
constituents by the $J_z={1\over 2}$ sum rule:
\begin{equation}
J_z={1\over 2}={1\over 2}\langle\Delta q_v\rangle+{1\over 2}\langle
\Delta S\rangle+\langle\Delta G\rangle+L_z. \label{11}
\end{equation}                                         
The right hand side represents the decomposition of the constituent spins along
with their relative angular momentum, $L_z$. Although this does not provide
a strict constraint on either $\Delta q_{tot}$ or $\Delta G$, it does give
an indication of the angular momentum component to proton spin.

\begin{center}  \Large
{\bf Phenomenology}
\end{center}   \normalsize

We use SMC$^{4}$, SLAC$^{5}$ and DESY$^{6}$ data to extract information
about the flavor dependence of the sea contributions to nucleon spin.
We can write the integrals of the polarized structure functions,
$\int_0^1 g_1^i\>dx$ in the terms of the axial-vector currents as:
$$
I^p\equiv \int_0^1 g_1^p(x) dx=\Biggl[{{A_3}\over {12}}+{{A_8}\over {36}}+
{{A_0}\over 9}\Biggr]\Bigl(1-\alpha_s^{corr}\Bigr),$$
\begin{equation}
I^n\equiv \int_0^1 g_1^n(x) dx=\Biggl[-{{A_3}\over {12}}+{{A_8}\over {36}}+
{{A_0}\over 9}\Biggr]\Bigl(1-\alpha_s^{corr}\Bigr), \label{12}
\end{equation}
$$
I^d\equiv (1-{3\over 2}\omega_D)\int_0^1 g_1^d(x) dx=\Biggl[{{A_8}\over {36}}
+{{A_0}\over 9}\Biggr]\Bigl(1-\alpha_s^{corr}\Bigr)(1-{3\over 2}
\omega_D),
$$
where $\omega_D$ is the probability that the deuteron will be in a D-state.
Using N-N potential calculations, the value of $\omega_D$ is about $0.058$.
$^{17}$ The BSR can then be used to extract an effective $I^p$ value from all
data using the form of equation (12) above.

Since the anomalous dimensions for the polarized distributions have an
additional factor of $x$ compared to the unpolarized case, early treatments of
the spin distributions assumed a form of: $\Delta q(x)\equiv xq(x)$ for all
flavors.$^{18}$ We have compared this form of the distributions to those
extracted from the recent data, using the defined ratio $\eta \equiv{{\langle
\Delta q_{sea}\rangle_{exp}}\over {\langle xq_{sea}\rangle_{calc}}}$ for each
flavor. Any deviation from $\eta=1$ would indicate that the simple model for
generating the polarized distributions is inaccurate.

In order togenerate the $x$-dependent distributions, we have used the
unpolarized distributions$^{12,13}$ with our extracted value of $\eta$ and the
assumption that: $\Delta q(x)\equiv \eta xq(x)$ for each of the sea flavors.
For the valence distributions, we have used equation (5) with the dilution
factor of reference 10. There is no reason {\it a priori} to suspect that
a global fit to the integrated distributions should imply a satisfactory
$x$-dependent fit to the data. However, our results indicate that this form
gives very good $x$-dependent parametrizations for the polarized distributions.

The analysis (for each polarized gluon model) proceeds as follows:
\begin{itemize}
\item
Extract a value of $I^p$ from either the data directly or via the
BSR in the form of equation (12),
\item
use equation (9) to extract $A_0$. Then the overall contribution to the quark
spin is found from $\langle \Delta q_{tot}\rangle=A_0+\Gamma$.
\item
Use the value $A_8$ from the hyperon data with equations (8) and (9) to extract
$\Delta s$ for the strange sea,
\item
find the total contribution from the sea from $\langle\Delta q_{tot}
\rangle=\langle\Delta q_v\rangle+\langle\Delta S\rangle$,
\item
determine $\epsilon$ and the distributions $\langle \Delta u\rangle_{sea}=
langle \Delta d\rangle_{sea}$ from equation (3) and the strange sea results.
\item
Finally, the $J_z$= 1/2 sum rule gives $L_z$.
\end{itemize}

Results for the integrated distributions are given in Table I.

\newpage
\begin{center} \large
{\bf Table I: Integrated Polarized Distributions: \\
$\Delta G=xG$ (above line), $\Delta G=0$ (below line)}
\end{center}  \normalsize

$$\begin{array}{cccccc}
  Quantity    &SMC(I^p)  &SMC(I^d)  &E154(I^n)  &E143(I^d) &HERMES \cr
              &          &          &           &          &(I^n)  \cr
  <\Delta u>_{sea} &-.077  &-.089   &-.063      &-.068    &-.050 \cr
  <\Delta s>       &-.037  &-.048   &-.020      &-.028    &-.010 \cr
  <\Delta u>_{tot} &0.85   &0.82    &0.87       &0.87     &0.90  \cr
  <\Delta d>_{tot} &-.42   &-.43    &-.39       &-.40     &-.36  \cr
  <\Delta s>_{tot} &-.07   &-.10    &-.04       &-.06     &-.02  \cr
  \eta_u = \eta_d  &-2.4   &-2.8    &-1.9       &-2.1     &-1.5 \cr
    \eta_s      &-2.0    &-3.0      &-1.2       &-1.6     &-0.6 \cr
   \epsilon     &1.09    &0.84      &2.10       &1.41     &4.00 \cr
    \Gamma      &0.06    &0.06      &0.08       &0.08     &0.07 \cr
     I^p        &0.136   &0.129     &0.134      &0.131    &0.135 \cr
  <\Delta q>_{tot} &0.36   &0.29    &0.45       &0.41     &0.52 \cr
  <\Delta G>    &0.46    &0.46      &0.45       &0.44     &0.44 \cr
     L_z        &-.14    &-.211      &-.18       &-.15     &-.22 \cr
 ------& -----& -----& ------& -----& ----- \cr
 <\Delta u>_{tot} & .83     & .80    & .85    & .84   & .88   \cr
 <\Delta d>_{tot} & -.44    & -.45   & -.41   & -.43  & -.39  \cr
 <\Delta s>_{tot} & -.09    & -.12   & -.07   & -.08  & -.04  \cr
     I^p          & .136    & .129   & .134   & .131  & .135  \cr
 <\Delta q>_{tot} & 0.30    & 0.23   & 0.37   & 0.33  & 0.45 \cr
  \Gamma          & 0.00    & 0.00   & 0.00   & 0.00  & 0.00 \cr
     L_z          & 0.35    & 0.39   & 0.32   & 0.35  & 0.28
\end{array}$$ 

From the results in Table I, it is obvious that the naive quark model is not
sufficient to explain the proton's spin characteristics. Nor is the simple
model for extracting the polarized distributions accurate. Thus, data have
indicated that we must modify our initial assumptions regarding constituent
contribution to proton spin. Some conclusions which can be drawn from the
data are: \\

(1) The total quark contribution to proton spin is between 1/4 and 1/2. The
errors in generating these results are due mostly to experimental errors and
determination of which model of the polarized gluons to use. Thus, the
uncertainties related to the quark spin content and the size of $\Delta G$ are
comparable. \\

(2) Considerable discussion regarding these measurements focuses on the
Ellis-Jaffe sum rule (EJSR),$^{3}$ which predicts the values of $g_1^p$
and $g_1^n$ using an unpolarized strange sea. This differs considerably with
the analysis of these data. The up and down sea contributions seem to agree
within a few percent. However, all of the proton and deuteron data imply a
larger polarized sea with the strange sea polarized greater than the positivity
bound.$^{19}$ It is interesting to note that the results obtained
from the SMC proton data are consistent with a recent lattice QCD calculation
of these parameters.$^{20}$ The results among the different experiments
preclude us from extracting specific contributions from each flavor to the
proton spin with any degree of certainty. However, the results from the various
data can be categorized into distinct models, characterized by the size of the
non-zero polarized sea. \\

(3) The values of $\eta$ deviate considerably from unity for most of the data,
implying that the relation between unpolarized and polarized distributions is
likely more complex than originally thought. \\

(4) This analysis implies that the anomaly correction is not large.
If the anomaly term were larger, due to a large $\Delta G$, the strange sea
would be positively polarized, while the other flavors are negatively
polarized. There is no known mechanism that would allow this cross polarization
of different flavors. Thus, we conclude that these data imply that the
polarized gluon distribution is of small to moderate size.
The key conclusions from the zero and small anomaly models are not significantly
different. Further, even if there are higher twist corrections to the anomaly
at small $Q^2$, the anomaly will not reconcile differences in the flavor
dependence of the polarized sea.$^{21}$ \\

(5) The orbital angular momentum extracted from data is also much smaller than
earlier values obtained from EMC data.$^{16}$ A point of interest is that
the zero $\Delta G$ model implies a positive orbital angular momentum, while
the other model gives a negative result. Thus, although $L_z$ is likely small,
its sign is still in question. \\

(6) The extracted value for $I^p$ is comparable for all data and well within
the experimental uncertainties. This implies agreement about the validity of
the Bjorken Sum Rule. We have arrived at this conclusion by using the BSR to
extract an effective $I^p$, in contrast to the experimental groups, which
used data to extract the BSR. There is general agreement that the BSR (and
thus QCD) is in tact. \\
 
Clearly, these experiments have contributed to the progress of understanding
the relative contributions of the constituents to the proton spin. They have
probed to smaller $x$ values, while decreasing the statistical and systematic
errors. This, coupled with theoretical progress in calculating higher order
QCD and higher twist corrections have allowed us to narrow the range of these
spin contributions. Although the flavor contributions to the proton spin cannot
be extracted precisely, the range of possibilities has been substantially
decreased. The main differences are the questions of the strange sea spin
content and the size of the polarized gluon distribution. We stress that more
experiments must be performed to determine the relative contributions from
gluons and various flavors of the sea.

The $x$-dependent distributions are in very good agreement with proton,
neutron and deuteron data. The graphs which follow compare $x$ binned data
with the polarized $x$-dependent distributions generated from results on
$\eta$ and equation (3). These were found using the GRV$^{12}$ and the
MRS$^{13}$ unpolarized distributions. The differences between the two sets of
distributions are at small-$x$, where the data is most uncertain.
More DIS experiments should be performed at small-$x$ to distinguish between
models and to address the controversy regarding which contributions to $g_1$ are
dominate in this kinematic region.

\begin{figure}
\epsfxsize=120mm
\centerline{\epsfbox{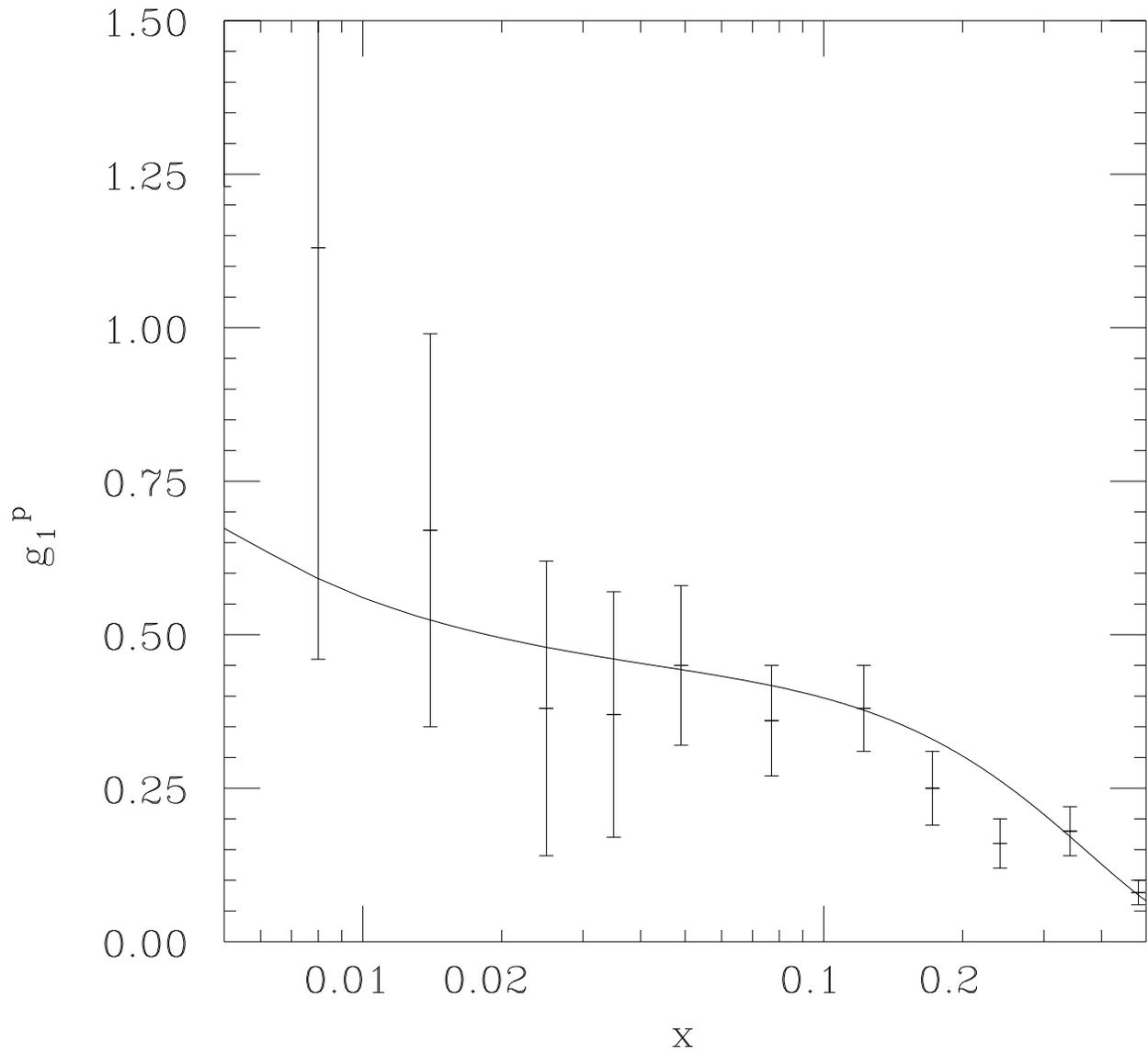}}
\caption{{\bf The $x$-dependent structure function $g_1^p$ is compared with
data as a function of $x$.}}
\end{figure}

\begin{figure}
\epsfxsize=120mm
\centerline{\epsfbox{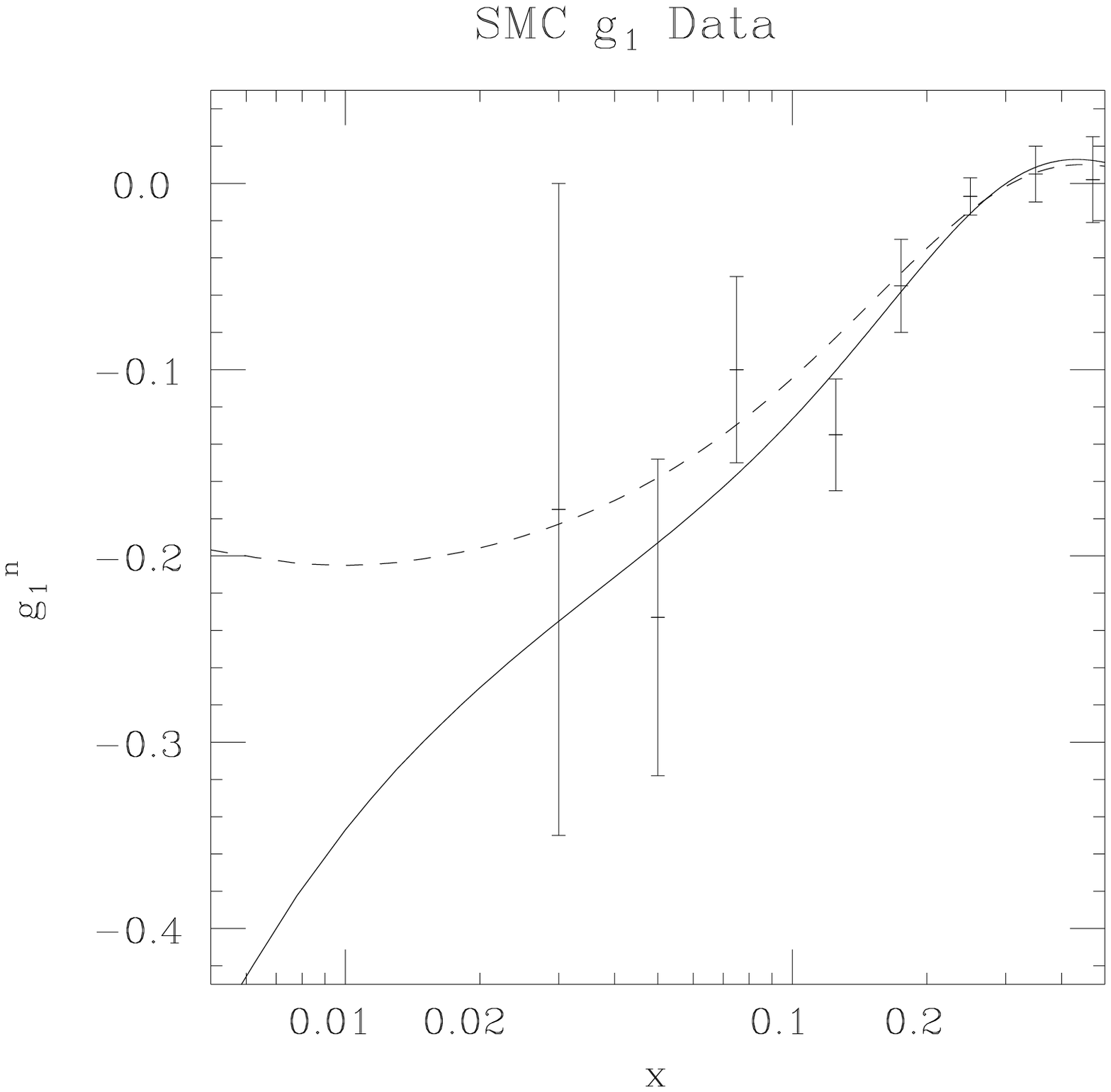}}
\caption{{\bf The $x$-dependent structure function $g_1^n$ is compared with
data as a function of $x$. The solid line represents the GRV generated
distributions and the dashed line the MRS generated distributions.}}
\end{figure}

\begin{figure}
\epsfxsize=120mm
\centerline{\epsfbox{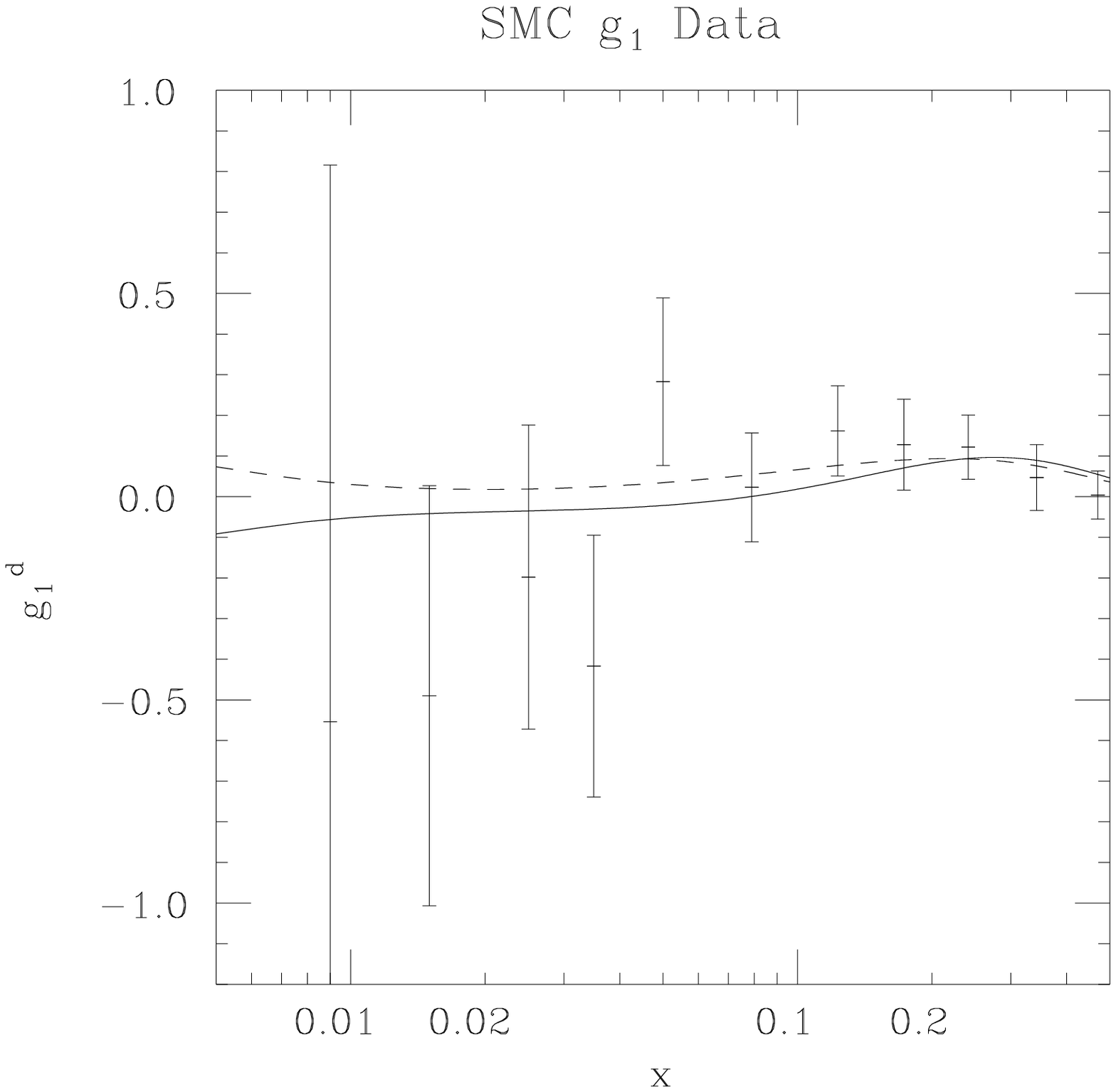}}
\caption{{\bf The $x$-dependent structure function $g_1^d$ is compared with
data as a function of $x$. The solid line represents the GRV generated
distributions and the dashed line the MRS generated distributions.}}
\end{figure}

\begin{center}  \Large
{\bf Possible Experiments}
\end{center}   \normalsize

There are a number of experiments which are technologically feasible that would
supply some of the missing information about these distributions. Detailed
summaries can be found in references 22 and 23. The large average luminosities
of these experiments and the success of Siberian Snakes makes all of the
following feasible.
 
Deep Inelastic Scattering: The E155 experiment has been approved at SLAC. These
experiments are designed to probe slightly smaller $x$ while greatly improving
statistics and systematical errors.
With lower error bars at small $x$, the extrapolation should achieve a more
accurate value for the integrated distributions and narrow the ranges of
constituent spin contributions even further.
   
Lepton Pair Production (Drell-Yan): 
The Relativistic Heavy Ion Collider (RHIC) at Brookhaven is designed so that
polarized $pp$ and $p\bar p$ experiments can be performed at large energy and
momentum transfer ranges. The energy range will be covered in discrete steps
of about 60, 250 and 500 GeV, but the momentum transfer range covers
$0.005 \le Q^2 \le 6.0$ GeV$^2$ in a fairly continuous set of steps.
The PHENIX detector is suitable for lepton detection and the wide range of
energies and momentum transfers could yield a wealth of Drell-Yan data over a
wide kinematic range. The x-dependence of the polarized sea distributions could
then be extracted to a fair degree of accuracy.

Jets, pions and direct photon production: 
The SPIN Collaboration proposes a set of experiments, which are in the
kinematic region where the measurement of double spin asymmetries in jet
production would give a sensitive test of the polarized gluon distribution's
size. Naturally, this measurement has an effect on both $\Delta G$
and the anomaly term appearing in the polarized quark distributions.$^{24}$
The STAR detector at RHIC is suitable for inclusive reactions involving jet
measurements, direct photon production$^{25}$ and pion production. All of these
would provide excellent measurements of the $Q^2$ dependence of $\Delta G$
since all are sensitive to the polarized gluon density at differing $Q^2$
values. Charm production in polarized collisions are also sensitive to $\Delta
G$ and should be performed at RHIC. Should DESY proceed with plans to polarize
their proton beam, many of these experiments could be performed there,
complementing the kinematic regions covered by RHIC and CERN. 
           
There has been considerable discussion about performing the COMPASS polarization
experiments at the LHC at CERN. Depending on the approved experiments,
there is the possibility of probing small $x$ and doing polarized
inclusive experiments to measure both sea and gluon contributions to proton
spin. These could be made in complementary kinematic regions to those of
the other accelerators. There are tentative plans to do polarized $W^{\pm}$
production, which provides a measure of the $x$-dependent sea distributions.
Polarized $W^{\pm}$ production is also planned at SLAC and would provide useful
sea information in a slightly different kinematic region than that of CERN.

Tests of the valence quark polarized distributions can be made, provided a
suitable polarized antiproton beam of sufficient intensity could be developed.
$^{26}$ This would provide a good test of the Bjorken sum rule via measurement
of $\langle \Delta q_v\rangle$ and the assumption of a flavor symmetric up and
down sea. This should be an experimental priority for the spin community.

Existing data indicate that the spin structure of nucleons is non-trivial and
has led to the formulation of a crucial set of questions to be answered about
this structure. The experiments discussed above can and should be performed in
order to shed light on the spin structure of the nucleons. \\

{\bf Acknowledgements} \\

A major portion of this work was done with Mehrdad Goshtasbpour of Shahid
Beheshti Univertsity in Tehran, Iran.
I would like to express my appreciation to the Local Organizing Committee and
in particular, Professor T. Morii for inviting me and providing support to
take part in this workshop.
\\

{\bf References} \\

\begin{enumerate}
\item\label{R1} J. Ashman, {\it et.al.}, Phys. Lett. {\bf B206}, 364, (1988);
Nucl. Phys. {\bf B328}, 1, (1989).
\item\label{R2} J.D. Bjorken, Phys. Rev. {\bf 148}, 1467, (1966).
\item\label{R3} J. Ellis and R.L. Jaffe, Phys. Rev. {\bf D9}, 3594, (1974).
\item\label{R4} B. Adeva, {\it et. al.}, Phys. Lett. {\bf B302}, 533, (1993) and
Phys. Lett. {\bf B320}, 400, (1994); D. Adams, {\it et. al.}, Phys. Lett. {\bf
B329}, 399, (1994).
\item\label{R5} P.L. Anthony, {\it et. al.}, Phys. Rev. Lett. {\bf 71}, 959,
(1993); K. Abe, {\it et. al.}, Phys. Rev. Lett. {\bf 74}, 346, (1995) and
hep-ex/9610007.
\item\label{R6} HERMES Collaboration: Presented at the XII International
Symposium on High Energy Spin Physics, Amsterdam, The Netherlands, Sept.
10-14, 1996.
\item\label{R7} D. deFlorian, {\it et. al.}, Phys. Rev. {\bf D51}, 37, (1995)
and J. Ellis and M. Karliner, Phys. Lett. {B341}, 397, (1995).
\item\label{R8} S.A. Larin, Phys. Lett. {\bf B334}, 192, (1994); S.A. Larin,
F.V. Tkachev and J.A.M. Vermaseren, Phys. Rev. Lett. {\bf 66}, 862, (1991).
\item\label{R9} F.E. Close and R.G. Roberts, Phys. Lett. {\bf B316}, 165,
(1993).
\item\label{R10} J.-W. Qiu, G.P. Ramsey, D.G. Richards and D. Sivers,
Phys. Rev, {\bf D41}, 65, (1990).
\item\label{R11} R. Carlitz and J. Kaur, Phys. Rev. Lett. {\bf 38}, 673, (1977).
\item\label{R12} M. Gluck, E. Reya and A. Vogt, Z. Phys. {\bf C48}, 471, (1990);
Z. Phys. {\bf C53}, 127, (1992); Phys. Lett. {\bf B306}, 391, (1993) and A.
Vogt, Phys. Lett. {\bf B354}, 145, (1995).
\item\label{R13} A.D. Martin, R.G. Roberts and W.J. Stirling, J. Phys. G
{\bf 19}, 1429, (1993); Phys. Rev. {\bf D47}, 867, (1993); Phys. Lett.
{\bf B306}, 145, (1993); Phys. Lett. {\bf B354}, 155, (1995); Phys. Rev.
{\bf D50}, 6734, (1994) and preprint RAL-95-021 and DTP/95/14 (1995).
\item\label{R14}  G. Ramsey, J.-W. Qiu, D.G. Richards and D. Sivers, Phys. Rev.
{\bf D39}, 361, (1989).
\item\label{R15} A.V. Efremov and O.V. Teryaev, JINR Report E2-88-287 (1988);
G. Alterelli and G.G. Ross, Phys. Lett. {\bf B212}, 391, (1988); R.D. Carlitz,
J.C. Collins, and A.H. Mueller, Phys. Lett. {\bf B214}, 229, (1988).
\item\label{R16} F.E. Close and R.G. Roberts, Phys. Rev. Lett. {\bf 60}, 1471,
(1988); S.J. Brodsky, J. Ellis and M. Karliner, Phys. Lett. {\bf B206}, 309,
(1988); M. Anselmino, B.L. Ioffe, and E. Leader Yad. Fiz. {\bf 49}, 214, (1989).
\item\label{R17} S. Platchkov, preprint DAPNIA SPhN 93 53: invited talk at the
14th Eur. Conf. on Few Body Problems, Amsterdam, 1993; M Lacombe, {\it et.
al.}, Phys. Lett. {\bf B101}, 139, (1981).
\item\label{R18} P. Chiapetta and J. Soffer, Phys. Rev. {\bf D31}, 1019, (1985);
M. Einhorn and J. Soffer, Nucl. Phys. {\bf B274}, 714, (1986).
\item\label{R19} G. Preparata, P.G. Ratcliffe and J. Soffer, Phys. Lett.
{\bf B273}, 306, (1991).
\item\label{R20} S.J. Dong, J.-F. Laga$\ddot{e}$ and K.F. Liu, preprint UK-01
(1995).
\item\label{R21} E. Stein, {\it et. al.}, Phys. Lett. {\bf B353}, 107, (1995).
\item\label{R22} G.P. Ramsey, Particle World, {\bf 4}, No. 3, 1995.
\item\label{R23} S.B. Nurushev, IHEP preprint IHEP 91-103, Protvino, Russia.
\item\label{R24} G.P. Ramsey, D. Richards and D. Sivers Phys. Rev. {\bf D37},
314, (1988).
\item\label{R25} E.L. Berger and J.-W. Qiu, Phys. Rev. {\bf D40}, 778, (1989);
P.M. Nadolsky, Z. Phys. {\bf C62}, 109, (1994).
\item\label{R26} G.P. Ramsey and D. Sivers, Phys. Rev. {\bf D43}, 2861, (1991).
\end{enumerate}
\end{document}